\documentclass[twocolumn,showpacs,showkeys,prd,superscriptaddress]{revtex4-2}

\usepackage{amsmath,amssymb,amsfonts,dsfont,mathrsfs,amsthm}
\usepackage{graphicx}
\usepackage{centernot}
\usepackage{hyperref}
\usepackage{xcolor}
\usepackage{comment}
\hypersetup{linktocpage,colorlinks=true,urlcolor=blue!80!red,linkcolor=blue,citecolor=red}
\usepackage{feynmf}
\usepackage{siunitx}
\usepackage{array}
\usepackage{ulem}
\usepackage{tikz}
\usepackage{braket}

\begin{document}

\title{Tunable BCS-BEC crossover, reentrant, and hidden quantum phase transitions in two-band superconductors with tunable valence and conduction bands}

\author{Giovanni \surname{Midei}}
\affiliation{School of Science and Technology, Physics Division, University of Camerino,
Via Madonna delle Carceri, 9B, 62032 - Camerino (MC), Italy}
\author{Andrea \surname{Perali}}
\affiliation{School of Pharmacy, Physics Unit, University of Camerino,
Via Madonna delle Carceri, 9B, 62032 - Camerino (MC), Italy}

\begin{abstract}
Two-band electronic structures with a valence and a conduction band separated by a tunable energy gap and with pairing of electrons in different channels can be relevant to investigate the properties of two-dimensional multiband superconductors and electron-hole superfluids, as monolayer FeSe, recently discovered superconducting bilayer graphene, and double-bilayer graphene electron-hole systems. This electronic configuration allows also to study the coexistence of superconductivity and charge density waves in connection with underdoped cuprates and transition metal dichalcogenides. By using a mean-field approach to study the system above mentioned, we have obtained numerical results for superconducting gaps, chemical potential, condensate fractions, coherence lengths, and superconducting mean-field critical temperature, considering a tunable band gap and different filling of the conduction band, for parametric choice of the pairing interactions. By tuning these quantities, the electrons redistribute among valence and conduction band in a complex way, leading to a new physics with respect to single-band superconductors, such as density induced and band-selective BCS-BEC crossover, quantum phase transitions, and hidden criticalities. At finite temperature, this phenomenon is also responsible for the non-monotonic behavior of the superconducting gaps resulting in a superconducting-normal state reentrant transition, without the need of disorder or magnetic effects.
\end{abstract}

\maketitle
\section{Introduction}
Multi-band and multi-gap superconductivity is a complex quantum coherent phenomenon with peculiar features that cannot be found in single-band and single-gap superconductors \cite{Milosevic2015}. The increased number of degrees of freedom in the condensate state allows for novel quantum effects which are unattainable otherwise, for instance enriching the physics of the BCS-BEC crossover \cite{Eagles1969, Leggett1980, Chen2005, Strinati2018}. Proximity to the crossover regime of the BCS-BEC crossover in multi-band superconductors having deep and shallow bands can determine a notable increase of superconducting gaps and critical temperature (T$_c$) \cite{Shanenko2010, Chen2012, Innocenti2010, Mazziotti2017}, associated with an higher mean-field T$_c$, together with optimal conditions for the screening of superconducting fluctuations \cite{Salasnich2019,Tajima2019a,Tajima2020}. Furthermore, the interplay of low-dimensional two-band systems allows for screening of fluctuations in systems composed by coupled quasi-2D bands or even in the vicinity of a van Hove singularity (e.g., in the case of quasi-1D), enabling shrinking of the pseudo-gap phase and robust high-critical temperatures \cite{Saraiva2020, Saraiva2021, Saraiva2022}.

Motivated by high temperature superconductivity and anomalous metallic state properties in underdoped cuprates, interest has grown in the pseudogap physics, in which a blurred gap persists in the normal state near the Fermi level. There are different models and explanations for this pseudogap, the simplest one being a smooth crossover from the BCS regime towards a Bose-Einstein condensation regime in which bound pairs form first at higher temperatures, and then below a critical temperature T$_c$ they condense, with the pseudogap being the excitation energy of the quasi-molecular pairs. Another explanation relevant for underdoped cuprates is the presence of other mechanisms different from pair fluctuations, such as charge density waves (CDWs) \cite{Gabovich2009, Gabovich2010, Arpaia2019, Perali1996} and their fluctuations that can modify the energy spectrum with opening of (pseudo)gaps and at the same time mediate Cooper pairing. Thus, systems in which CDWs and superconductivity coexist are of primary interest to study the BCS-BEC crossover when an energy gap separates the electronic spectrum in two bands, determining a valence and a conduction band.

In addition to underdoped cuprates, an interesting example is given by the transition metal dichalcogenide (TMD) family, MX$_2$, where M = Ti, Nb, Mo, Ta and X = S, Se, which exhibits a rich interplay between superconductivity and CDW order \cite{Rossnagel2011}. In these materials, superconductivity occurs in an environment of pre-existing CDW order \cite{Neto2001,Kiss2007}, making them an ideal platform to study many-body ground states and competing phases in the 2D regime. The relationship between CDW and superconductivity in such systems is still under investigation \cite{Calandra2009, Ge2012}. In general, their mutual interaction is competitive, but evidence to the contrary, indicating a cooperative interplay, has also been reported in angle-resolved photoemission spectroscopy (ARPES) studies \cite{Kiss2007}. 
Among them, bulk Niobium diselenide (2H-NbSe$_2$) undergoes a CDW distortion at T=$30$ K and becomes superconducting at $7$ K. References \cite{Ugeda2016, Cao2015} reported that T$_c$ lowers to $1.9$ K in 2H-NbSe$_2$ single-layers and that the CDW measured in the bulk is preserved. Theoretical support is given by Chao-Sheng Lian et al. \cite{Lian2019}: they demonstrate enhanced superconductivity in the CDW state of monolayer tantalium diselenide (TaSe$_2$) with DFT calculations. In contrast with 2H-NbSe$_2$, they report that as TaSe$_2$ is thinned to the monolayer limit, its superconducting critical temperature rises from $0.14$ K in the bulk to $2$ K in the monolayer.
Another appealing superconducting material is the monolayer FeSe grown on a SrTiO$_3$ substrate, which exhibits a huge increase of T$_c$ up to $100$ K \cite{Liu2015} and it is characterized by a valence and a conduction band structure near the Fermi level.
Furthermore, very recently 2D superconductivity has been found in bilayer graphene systems, in which conduction and valence bands are separated by a small energy band-gap ($0\div100$ meV) that can be 
precisely tuned by an external electric field \cite{Zhou2022} (for a review see \cite{Pantaleon2022}). Coupling a monolayer of WSe$_2$ with bilayer graphene has been found to enhance superconductivity by an order of magnitude in T$_c$ and superconductivity emerges already at zero magnetic field \cite{Zhang2022}. 
Finally, it turns out that the two-band superconducting system considered in this work is in close correspondence with two-band electron-hole superfluids in double bilayer graphene \cite{Conti2017}.\par Therefore, the growing experimental realization of 2D superconductors with valence and conduction bands separated by a tunable energy gap and electron-hole superfluidity in multilayer systems motivated us to investigate the BCS-BEC crossover in this kind of systems. The detailed analysis of this configuration is lacking in the literature to the best of our knowledge. A pioneering work on a related system with valence and conduction parabolic bands has been done by Nozières and Pistolesi \cite{Nozieres1999} to study the phase transition from a semiconducting to a superconducting state and the consequent (pseudo)gap opening, in the specific case of equal pairing strengths for all interaction channels considered. In our work we consider a superconductor with two tight-binding bands with different intra-band and pair-exchange couplings, in order to probe the possibility to have coexisting Cooper pairs of different average sizes \cite{Tajima2019b} in the valence and conduction band. However, for most of multi-band superconductors the tuning of intra-band and pair-exchange interactions is rather challenging and their properties cannot be studied easily in a continuous way across the BCS-BEC crossover. As shown in this work, a different way to explore the BCS-BEC crossover in such systems can be achieved by tuning the energy gap between the valence and the conduction band. In fact, since the number of particles in the single bands is not conserved, when the energy band gap is modified the number of holes and of electrons forming Cooper pair respectively in the valence and in the conduction bands changes, allowing for the occurrence of a density induced multi-band BCS-BEC crossover \cite{Andrenacci1999}. This redistribution of charges between the valence and the conduction band leads also to novel and interesting quantum phase transitions (QPTs) from a superconducting to an insulating state, or hidden criticalities evidenced by the analysis of the order parameter coherence lengths \cite{Ord2012, Carlos2022}. At finite temperature, a new type of reentrant superconducting to normal state transition has been also found and characterized. The results reported and discussed in this work demonstrate the richness of the proposed valence and conduction band configuration to generate and tune new types of crossover phenomena and quantum phases.

The manuscript is organized as follow. In section II we describe the model for the physical system considered and the theoretical approach for the evaluation of the superconducting state properties. In section III we report our results. The conclusions of our work will be reported in Section IV.   

\section{Model system and theoretical approach}
\begin{figure}[b]
\includegraphics[width=.47\textwidth]{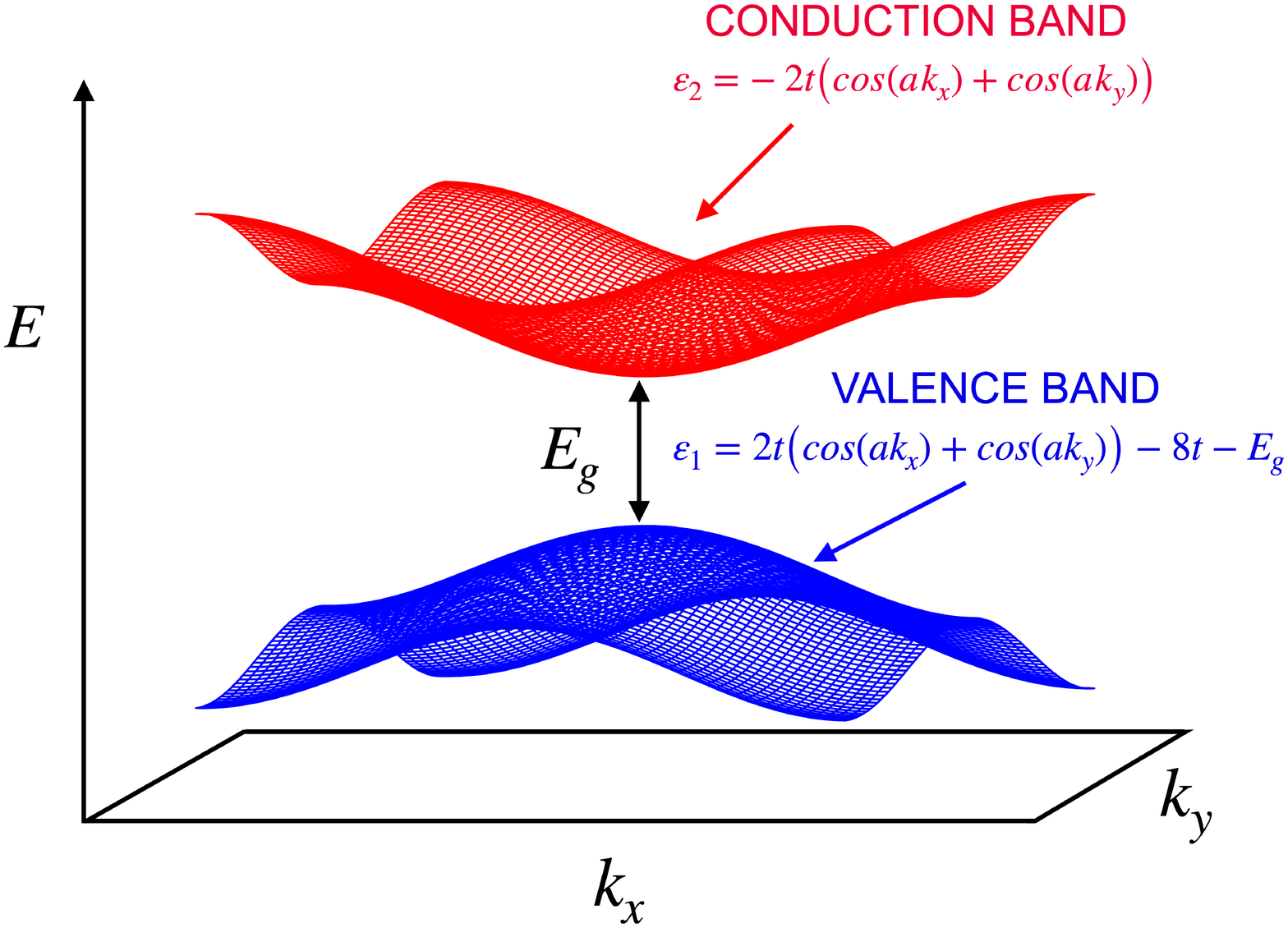}
\caption{Electronic band structure of the two-band 2D system considered in this work. $E_g$ is the energy gap between the valence (i = 1) and the conduction (i = 2) band.}
\label{fig1}
\end{figure}

We consider a two-dimensional (2D) two-band superconductor with a valence and a conduction electronic band in a square lattice. The valence and the conduction bands are modelled by a tight-binding dispersion given, respectively, by Eqs. (\ref{eqn:11}) and (\ref{eqn:12}): 
\begin{equation}
\varepsilon_1(\mathbf{k})=2t[\cos(k_xa)+\cos(k_ya)]-8t-E_g
\label{eqn:11}
\end{equation}
\begin{equation}
\varepsilon_2(\mathbf{k})=-2t[\cos(k_xa)+\cos(k_ya)]
\label{eqn:12} 
\end{equation}
where $t$ is the nearest neighbour hopping parameter assumed to be the same for both bands, $a$ is the lattice parameter and the wave-vectors belong to the first Brillouin zone $-\frac{\pi}{a} \leq k_{x,y} \leq \frac{\pi}{a}$; $E_g$ is the energy band-gap between the conduction and the valence band. The band dispersions are reported in Fig. \ref{fig1}.
In order to study the superconducting state properties of our system, we assume that Cooper pairs formation is due to an attractive interaction between opposite spin electrons. The two-particle interaction has been approximated by a separable potential $V_{ij}(\mathbf{k}, \mathbf{k'})$ with an energy cutoff $\omega_0$, which is given by:
\begin{equation}
V_{ij}(\mathbf{k}, \mathbf{k'})= -V_{ij}^{0} \Theta\Bigl(\omega_0-|\xi_i(\mathbf{k})|\Bigr) \Theta\Bigl(\omega_0-|\xi_i(\mathbf{k'})|\Bigr)
\label{eqn:13}
\end{equation}
where $V_{ij}^0>0$ is the strength of the potential in the different pairing channels and $i,j$ label the bands. $V_{11}^0$ and $V_{22}^0$ are the strength of the intra-band pairing interactions (Cooper pairs are created and destroyed in the same band).
$V_{12}^0$ and $V_{21}^0$ are the strength of the pair-exchange interactions (Cooper pairs are created in one band and destroyed in the other band, and vice versa), so that superconductivity in one band can induce superconductivity in the other band.
The same energy cutoff $\omega_0$ of the interaction for intra-band and pair-exchange terms is considered. 
Through out this work, $\omega_0$ is considered an energy scale larger than the total bandwidth of our system to model an effective pairing of electronic origin,
or a contact attractive potential. This is a key assumption to make possible for the system to explore the entire BCS-BEC crossover \cite{Guidini2014}.
The terms corresponding to Cooper pairs forming from electrons associated with different bands (inter-band or cross-band pairing) are not considered in this work (see \cite{Paredes2020}).
$\xi_i(\mathbf{k})=\varepsilon_i(\mathbf{k})-\mu$ in Eq. (\ref{eqn:13}) is the energy dispersion for the band $i$ with respect to the chemical potential $\mu$.
The superconducting state of the system and its evolution with relevant system parameters is studied at a mean-field level. The BCS equations for the superconducting gaps have to be coupled with the density equation which fixes the chemical potential, since the self-consistent renormalization of the chemical potential is a key feature to account for the BCS-BEC crossover physics. Zero and finite temperature cases have been considered in this work.
The BCS equations for the superconducting gaps in the two-band system at a given temperature T are
\begin{equation}\begin{split}
\Delta_1(\mathbf{k})=&-\frac{1}{2\Omega} \sum_{k'}\Biggl [V_{11}(\mathbf{k}, \mathbf{k'})\frac{\Delta_1(\mathbf{k'})}{E_1(\mathbf{k'})} \tanh{\frac{E_1(\mathbf{k'})}{2 T}}\\
&+V_{12}(\mathbf{k}, \mathbf{k'})\frac{\Delta_2(\mathbf{k'})}{E_2(\mathbf{k'})} \tanh{\frac{E_2(\mathbf{k'})}{2 T}}\Biggr]
\label{eqn:14}
\end{split}\end{equation}
\begin{equation}\begin{split}
\Delta_2(\mathbf{k})=&-\frac{1}{2\Omega} \sum_{k'}\Biggl [V_{22}(\mathbf{k}, \mathbf{k'})\frac{\Delta_2(\mathbf{k'})}{E_2(\mathbf{k'})} \tanh{\frac{E_2(\mathbf{k'})}{2 T}}\\
&+V_{21}(\mathbf{k}, \mathbf{k'})\frac{\Delta_1(\mathbf{k'})}{E_1(\mathbf{k'})} \tanh{\frac{E_1(\mathbf{k'})}{2 T}}\Biggr]
\label{eqn:15}
\end{split}\end{equation}
where $E_i(\mathbf{k})=\sqrt{{\xi_i(\mathbf{k})}^2+{\Delta_i(\mathbf{k})}^2}$ is the dispersion of single-particle excitations in the superconducting state and $\Omega$ is the area occupied by the 2D system. $\hbar=1$ and $k_B=1$ throughout the manuscript. The superconducting gaps have the same energy cutoff of the separable interaction:
\begin{equation}
\Delta_i(\mathbf{k})=\Delta_i \Theta\Bigl(\omega_0-|\xi_i(\mathbf{k})|\Bigr)
\label{eqn:16}
\end{equation}
The total electron density of the two-band system is fixed and given by the sum of the single-band densities, $n_{tot} = n_1 + n_2$, that can vary instead. The electronic density $n_i$ in the band ($i$) at temperature T is given by,
\begin{equation}
n_i=\frac{2}{\Omega}  \sum_{k} \Bigl[{v_i(\mathbf{k})}^2 f\big(-E_i(\mathbf{k})\big)+{u_i(\mathbf{k})}^2 f\big(E_i(\mathbf{k})\big)\Bigr]
\label{eqn:17}
\end{equation}
where $f(E)$ is the Fermi-Dirac distribution function. The BCS coherence weights $v_i(\mathbf{k})$ and $u_i(\mathbf{k})$ are:
\begin{equation}
{v_i(\mathbf{k})}^2=\frac{1}{2}\Bigg[1- \frac{\xi_i(\mathbf{k})}{\sqrt{{\xi_i(\mathbf{k})}^2+{\Delta_i(\mathbf{k})}^2}}\Bigg]
\label{eqn:18}
\end{equation}
\vspace{.2cm}
\begin{equation}
{u_i(\mathbf{k})}^2=1-{v_i(\mathbf{k})}^2
\label{eqn:19}
\end{equation}
For the valence band the definition of the condensate fraction is the ratio of the number of Cooper pairs in the valence band to the number of holes in the valence band,
\begin{equation}
\alpha^h_1=\frac{ \sum_{\mathbf{k}} {\bigl(u_1(\mathbf{k}) v_1(\mathbf{k})}\bigr)^2}{\sum_{\mathbf{k}} {u_1(\mathbf{k})}^2}
\label{eqn:603}
\end{equation}
For the conduction band instead, the expression already used in the one-band case is generalized to the number of Cooper pairs divided by the total number of carriers in the conduction band
\begin{equation}
\alpha^e_2=\frac{ \sum_{\mathbf{k}} {\bigl(u_2(\mathbf{k}) v_2(\mathbf{k})}\bigr)^2}{\sum_{\mathbf{k}} {v_2(\mathbf{k})}^2}
\label{eqn:604}
\end{equation}
The intra-pair coherence length $\xi_{pair_i}$ has the same form for both the valence and the conduction bands, that is
\begin{equation}
\xi^2_{pair_i}=\frac{ \sum_{\mathbf{k}} {\big|\nabla \bigl(u_i(\mathbf{k}) v_i(\mathbf{k})\bigr)}\big|^2}{\sum_{\mathbf{k}} {\bigl(u_i(\mathbf{k}) v_i(\mathbf{k})}\bigr)^2}
\label{eqn:605}
\end{equation}
Regarding the superconducting order parameter coherence length, two characteristic length scales in the spatial behavior of superconducting fluctuations are expected, since the system is made up by two partial condensates. When the pair-exchange interaction is not present, these two lengths are simply the order parameter coherence lengths of the condensates of the valence $\xi_{c 1}$ and of the conduction $\xi_{c 2}$ band. When the pair-exchange interactions is different from zero, one has to deal with coupled condensates, and these length scales cannot be attributed to the single bands involved, describing instead the collective features of the whole two-component condensate. The pair-exchange interactions mix the superconducting order parameters of the initially non-interacting bands, that acquire mixed character. The soft, or critical, coherence length $\xi_s$ diverges at the phase transition point, while the rigid, or non-critical, coherence length $\xi_r$ remains finite. Following the approach in \cite{Ord2012}, these characteristic length scales are given by 
\begin{equation}
\xi^2_{s, r}= \frac{G(T) \pm \sqrt{G^2(T)-4K(T) \gamma(T)}}{2 K(T)}
\label{eqn:607}
\end{equation}
where $\xi_s$ corresponds to the solution with the plus and $\xi_r$ to the one with the minus sign and
\begin{equation}
\begin{split}
G(T)=(V_{12}^0)^2\bigl(\tilde{g}_1(T)\beta_2(T)+\tilde{g}_2(T)\beta_1(T)\bigr)+\\
\bigl(1-V_{11}^0 \tilde{g}_1(T)\bigr)V_{22}^0 \beta_2(T)+\\
\bigl(1-V_{22}^0 \tilde{g}_2(T)\bigr)V_{11}^0 \beta_1(T)
\label{eqn:608}
\end{split}
\end{equation}

\begin{equation}
\begin{split}
K(T)=\bigl(1-V_{11}^0 \tilde{g}_1(T)\bigr)\bigl(1-V_{22}^0 \tilde{g}_2(T)\bigr)-\\
(V_{12}^0)^2 \tilde{g}_1(T)\tilde{g}_2(T)
\label{eqn:609}
\end{split}
\end{equation}

\begin{equation}
\gamma(T)=\bigl(V_{11}^0 V_{22}^0-(V_{12}^0)^2\bigr)\beta_1(T)\beta_2(T)
\label{eqn:610}
\end{equation}

\begin{equation}
\tilde{g}_i(T)=g_i(T)-3\nu_i(T)\bigl(\Delta_i(T)\bigr)^2
\label{eqn:611}
\end{equation}

\begin{equation}
g_i(T)= \frac{1}{2V} \sum_\mathbf{k} \frac{1}{\xi_i(\mathbf{k})} \tanh{\frac{\xi_i(\mathbf{k})}{2 T}}
\label{eqn:612}
\end{equation}

\begin{equation}
\begin{split}
\nu_i&(T)=\\
&-\frac{1}{2V} \sum_\mathbf{k} \frac{\partial}{\partial |\Delta_i|^2} \Biggl(\frac{1}{E_i(\mathbf{k})}  \tanh{\frac{\xi_i(\mathbf{k})}{2 T}}\Biggr)_{\Delta_i=0}
\label{eqn:613}
\end{split}
\end{equation}

\begin{equation}
\begin{split}
\beta_i&(T)=-\frac{1}{4V} \sum_\mathbf{k} \frac{\partial^2} {\partial q^2_l} \Biggl[\frac{1}{\xi_i(\mathbf{k})+\xi_i(\mathbf{k}-\mathbf{q})}\\
&\times \Bigl(\tanh{\frac{\xi_i(\mathbf{k})}{2 T}}+\tanh{\frac{\xi_i(\mathbf{k}-\mathbf{q})}{2 T}}\Bigr)\Biggr]_{\mathbf{q}=0}
\label{eqn:614}
\end{split}
\end{equation}
where $l$ refers to the Cartesian axis in Eq. (\ref{eqn:614}).

In order to describe the physics of the quantum phase transition, the values of the coherence lengths at zero temperature have been approximated by choosing a low enough temperature so that the superconducting gaps and the chemical potential retain the same behavior of the zero temperature case.
The energies are normalized in units of the hopping $t$ and the dimensionless couplings $\lambda_{ii}$ are defined as $\lambda_{ii}=N V_{ii}^0$, where $N=1/4 \pi a^2 t$ is the density of states at the top / bottom of the valence / conduction band, that coincide since the density of states is not modified by the concavity of the band.
The intra-pair coherence lengths $\xi_{pair_i}$ are normalized using the average inter-particle distance in the normal state $l_i=1/ \sqrt{\pi n_i}$, where $n_i$ is the density in the band $i$. This quantities differ by a factor of $\sqrt{2}$ by the inverse of the respective Fermi wave-vector $K_{F i}$.
The soft $\xi_s$ and the rigid $\xi_r$ coherence lengths are normalized with respect to the lattice constant $a$, since in the two-band case they cannot be attributed to any of the two bands.

\section{Results}
\begin{figure}[h]
\hspace{-0.3cm}\includegraphics[width=.49\textwidth]{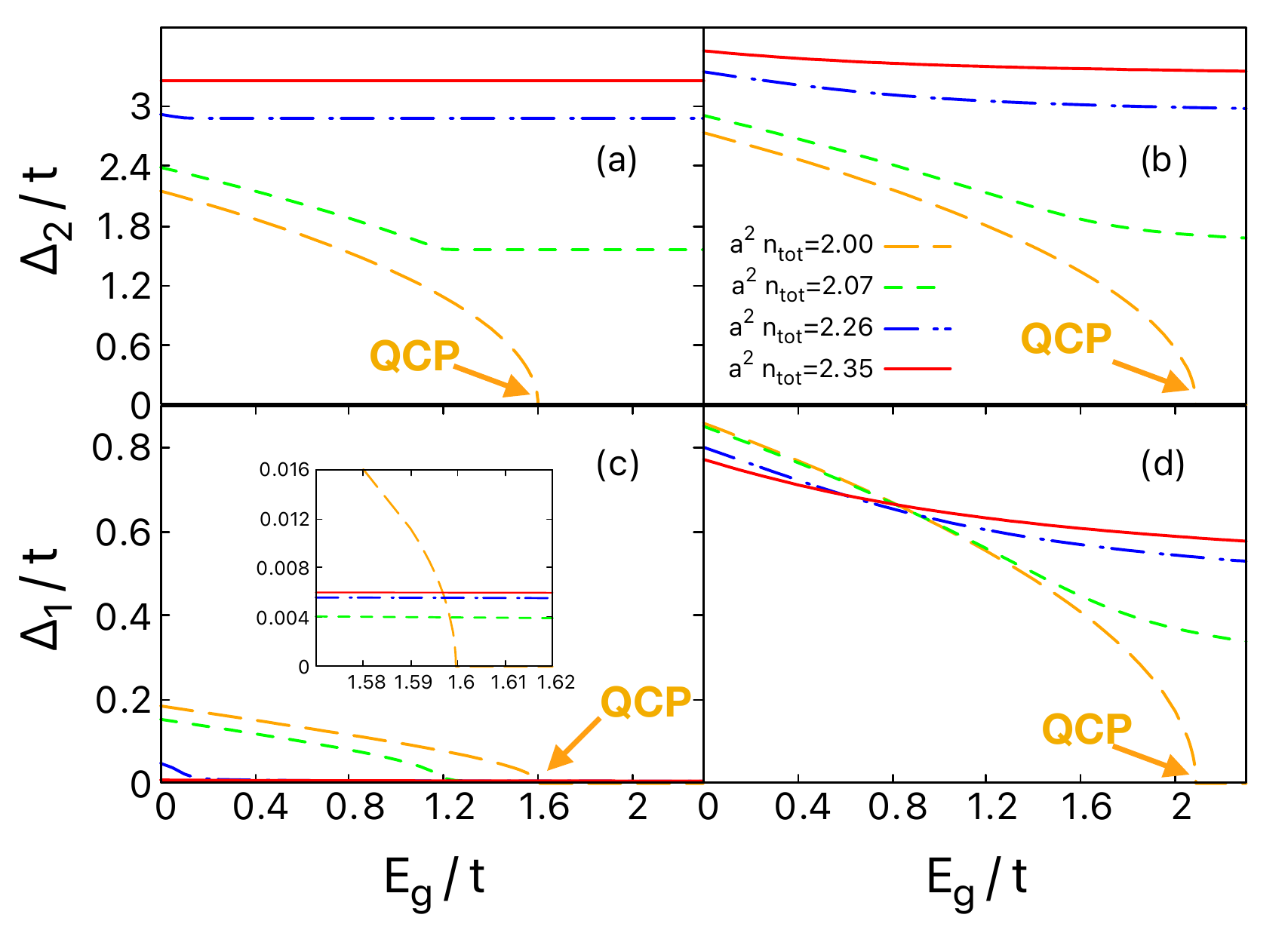}
\caption{Superconducting gaps $\Delta_2 / t$ opening in the conduction band (a)-(b) and in the valence band $\Delta_1 / t$ (c)-(d) as functions of the band-gap energy $E_g / t$ for an energy cutoff of the attractive interactions $\omega_0 / t=20$. The intra-band couplings are $\lambda_{11}=0.23$ and $\lambda_{22}=0.75$. The pair-exchange couplings are ($\lambda_{12}=\lambda_{21}$): (a),(c) (0.001), (b),(d) (0.1). The superconducting gaps are reported for different values of the total density $a^2 n_{tot}$.}
\label{fig2}
\end{figure}

In this section we study the properties of the superconducting ground state and give a full characterization of the BCS–BEC crossover in the two-band system considered in this work. First, we study the zero temperature superconducting gaps in the conduction ($\Delta_2$) and in the valence ($\Delta_1$) band through the BCS-BEC crossover, for the case of unbalanced intra-band couplings ($\lambda_{11} \neq \lambda_{22}$). The results are shown in Fig. \ref{fig2}, in which the superconducting gaps are reported as functions of the energy band-gap $E_g$, for different values of the total density $a^2 n_{tot}$ and for different pair-exchange couplings $\lambda_{12} = \lambda_{21}$.
 In the case of an empty conduction band and a completely filled valence band, corresponding to $a^2 n_{tot}=2.00$, a quantum phase transition (QPT) to the normal state takes place at a specific quantum critical point (QCP), that occurs when $E_g=E_g^*$. When the carrier concentration in the conduction band is non-zero, the phase transition becomes a crossover and superconductivity extends for all values of the band gap $E_g$. However, the system presents different behaviors if the value of the band gap is smaller or larger of $E_g^*$. For finite doping, the valence band contributes very weakly to the superconducting state of the system for  $E_g>E_g^*$. In this regime the bands are almost decoupled and the superconducting gaps does not depend on $E_g$. However, in the case of Fig. \ref{fig2}(c) since the pair-exchange couplings are weak the conduction band cannot sustain the superconductivity in the valence band and $\Delta_1$ is suppressed. Thus, continuously tuning $E_g$ to higher values will result in $\Delta_1<< \Delta_2$ so that there is only one significant superconducting gap and one significant condensate. In the other case instead (Fig. \ref{fig2}(d)), the pair-exchange couplings are stronger and $\Delta_1$ is not much suppressed with respect to its initial value, since in these cases the superconductivity in the valence band is sustained by the condensate of the conduction band. 
\\Another interesting feature of this system is that $\Delta_1$ is enhanced for lower values of the total density as long as $E_g<E_g^*$. When $E_g>E_g^*$ instead, the opposite situation occurs. The value of $E_g^*$ at which this behavior takes place depends on the level of filling of the conduction band, shifting to the left when higher total densities are considered, and on the pair-exchange couplings that shifts $E_g^*$ to the right when larger interactions strength are considered.
\begin{figure}[h]
\hspace{-0.4cm}\includegraphics[width=.5\textwidth]{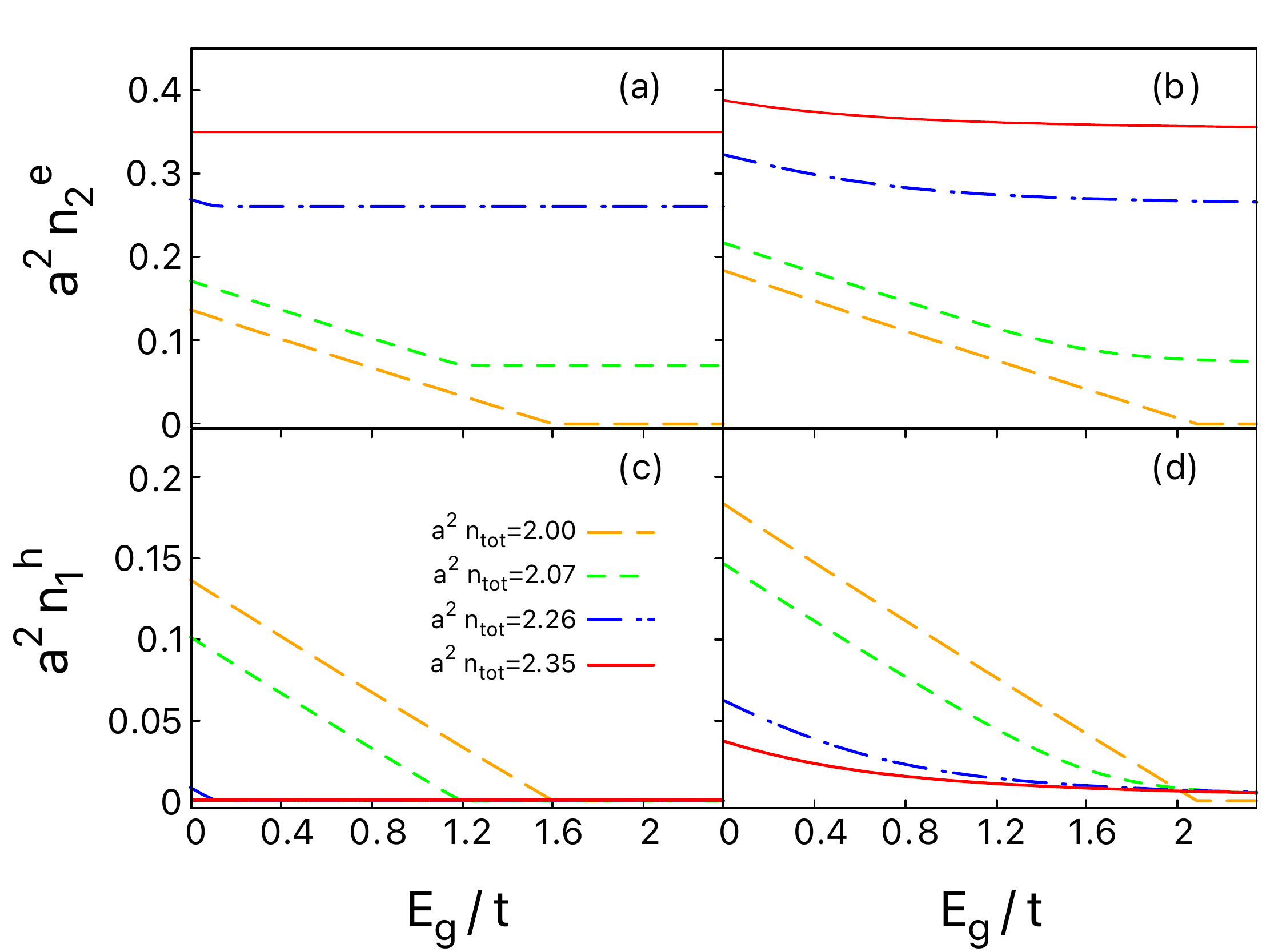}
\caption{Electron density $a^2 n_2^e$ (a)-(b) in the conduction band and hole density $a^2 n_1^h$ (c)-(d) in the valence band as functions of the band-gap $E_g/t$ for different values of the total density $a^2 n_{tot}$, normalized to the area of the unit cell. $\omega_0 / t=20$. The intra-band couplings are $\lambda_{11}=0.23$ and $\lambda_{22}=0.75$. The pair-exchange couplings are ($\lambda_{12}=\lambda_{21}$): (a),(c) (0.001), (b),(d) (0.1).}
\label{fig3}
\end{figure}
The reason behind the behavior of the superconducting gaps can be found by looking at the densities of particles forming Cooper pairs, which are electrons in the conduction band and holes in the valence band. 
While the total density is fixed, the density in each band can vary. In this way, the density of particles in the conduction band $n_2$ is no longer controlled only by doping as for a single band system, there are instead additional particles excited from the valence band. Nevertheless, for larger values of $E_g$ the gain in the interaction energy due to superconductivity is much smaller than the kinetic energy cost for transferring electrons from the valence band to the conduction band, so that very few electrons (compared to the total density of electrons in the valence band) are excited into the conduction band. This behavior is shown in Fig. \ref{fig3}. As one can see for $a^2 n_{tot} = 2.00$ the hole density in the valence band and the electron density in the conduction band coincide and are monotonically decreasing, both of them vanishing at the QCP $E_g=E_g^*$. This is a sign that superconductivity is due to holes in the valence band and to electrons in the conduction band. In the other cases the hole density in the valence band is almost zero for $E_g>E_g^*$, while the electron density in the conduction band is approaching the asymptotic value given by the total density minus the density of the filled valence band  $a^2 n_2= a^2 n_{tot}-2.00$. 
\begin{figure}[h]
\hspace{-0.cm}\includegraphics[width=0.5\textwidth]{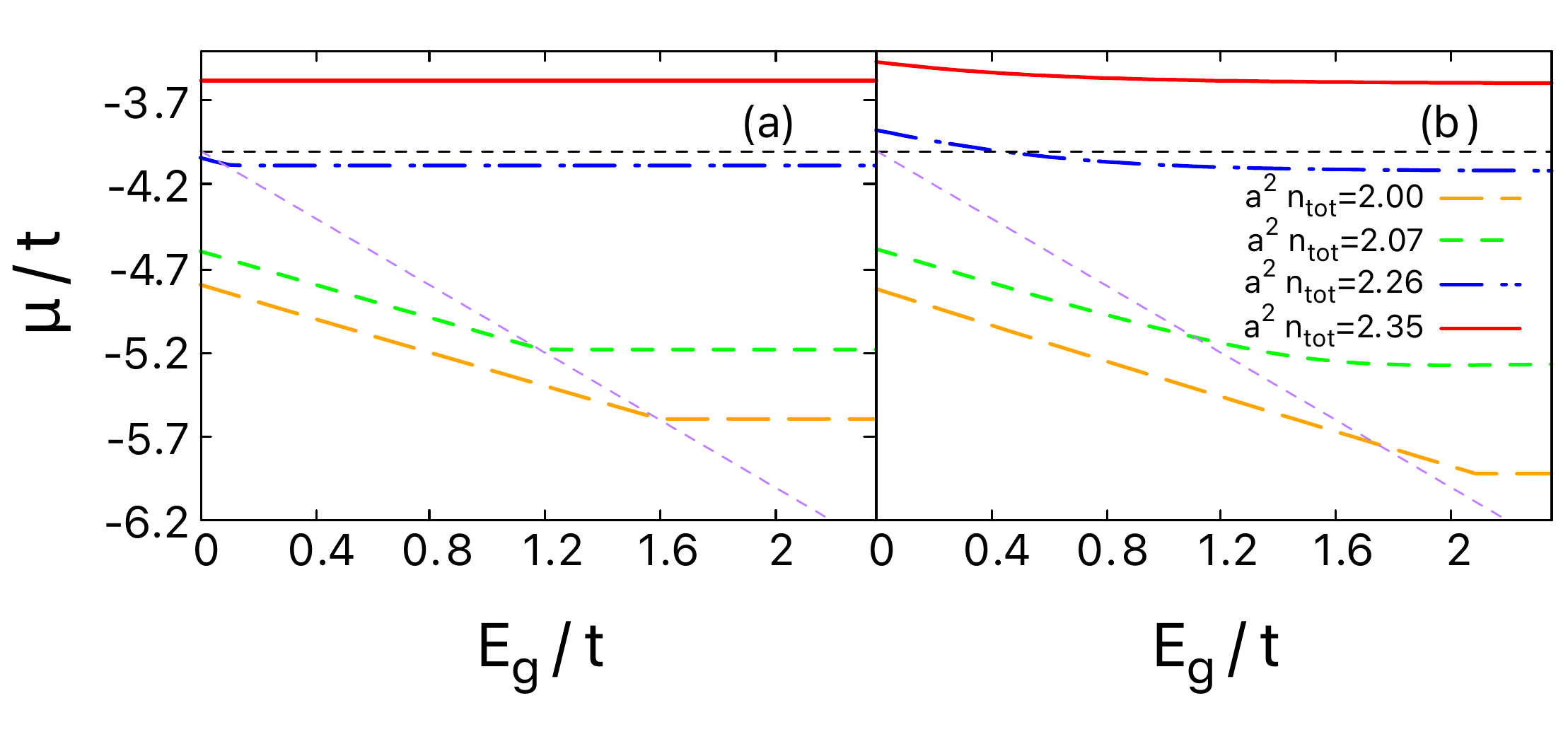}
\caption{Chemical potential $\mu / t$ as a function of the band-gap $E_g / t$ for $\omega_0 / t=20$. The pair-exchange couplings are $\lambda_{11}=0.23$ and $\lambda_{22}=0.75$. The pair-exchange couplings are ($\lambda_{12}=\lambda_{21}$): (a) (0.001),(b) (0.1). The chemical potential $\mu$ is reported for different total densities $a^2 n_{tot}$. The black and the magenta dashed lines correspond to the bottom of the conduction band and the top of the valence band, respectively.}
\label{fig4}
\end{figure}
\\In Fig. \ref{fig4} the chemical potential is reported as a function of $E_g$, for different total densities $a^2 n_{tot}$ and for different pair-exchange couplings.
For higher values of the total density and of the pair-exchange couplings the chemical potential shift toward higher energies, due to the larger number of electrons in the conduction band. In particular, when $E_g$ is increased, in the low density regime the chemical potential starts deep inside the valence band and then enters the gap between the two bands, meaning that the condensate in the valence band spans a wide region of the BCS-BEC crossover, while the conduction band is always located in the BEC side of the crossover regime or in the BEC regime, depending on whether the chemical potential lies inside the conduction band or not. 
When $E_g>E_g^*$ the chemical potential acquires a flat dependence and is not modified by $E_g$, in a similar way to what happens to the superconducting gaps and the densities. 
\begin{figure}
\hspace{-0.7cm}\includegraphics[width=0.50\textwidth]{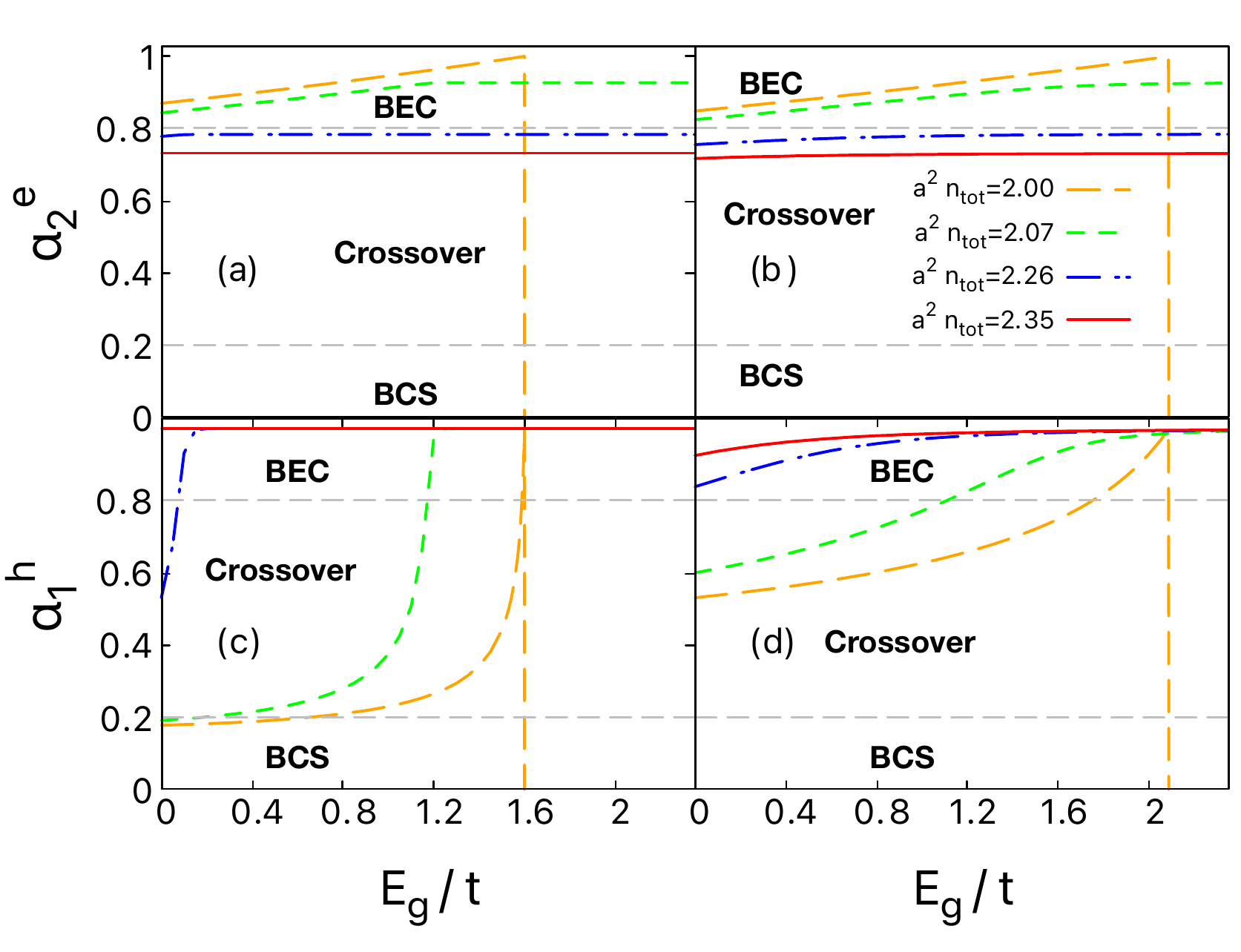}
\caption{Condensate fractions in the conduction band $\alpha^e_2$ (a)-(b) and in the valence band $\alpha^h_1$ (c)-(d) as functions of the band-gap $E_g/t$ for $\omega_0 / t=20$. The intra-band couplings are $\lambda_{11}=0.23$ and $\lambda_{22}=0.75$. The pair-exchange couplings are ($\lambda_{12}=\lambda_{21}$): (a),(c) (0.001), (b),(d) (0.1). The condensate fractions are reported for different total densities $a^2 n_{tot}$. Thin grey dashed lines correspond to $\alpha = 0.2, 0.8$ from bottom to top.}
\label{fig5}
\end{figure}
\\In Fig. \ref{fig5} the condensate fraction is shown as a function of $E_g$, for different $a^2 n_{tot}$ and for different pair-exchange couplings. The usual choice of the boundaries between the different pairing regimes has been adopted: for $\alpha < 0.2$ the superconducting state is in the weak-coupling BCS regime; for $0.2 < \alpha < 0.8$ the system is in the crossover regime; for $\alpha > 0.8$ the system is in the strong-coupling BEC regime.
Consistently with the information obtained from the chemical potential, in the low density regime the condensate in the valence band explores the entire BCS-BEC crossover by varying $E_g$. For the considered pair-exchange interactions in (Fig. \ref{fig5}(c)) the valence band condensate is in the BCS regime for small $E_g$, while for larger pair-exchange interactions (Fig. \ref{fig5}(d)) is in the crossover regime. When the energy gap or the total density increases, the valence band condensate enters the BEC regime, with the hole condensate fraction $\alpha^h_1$ approaching unity, indicating that the remaining few holes are all in the condensate. The situation in the conduction band is different, since due to the strong intra-band coupling the condensate is always located in the BEC side of the crossover regime or in the BEC regime. 
In the case $a^2 n_{tot}=2.00$ both the condensate fractions suddenly drop to zero when $E_g=E_g^*$ due to the quantum phase transition.
\begin{figure}
\hspace{-0.5cm}\includegraphics[width=0.5\textwidth]{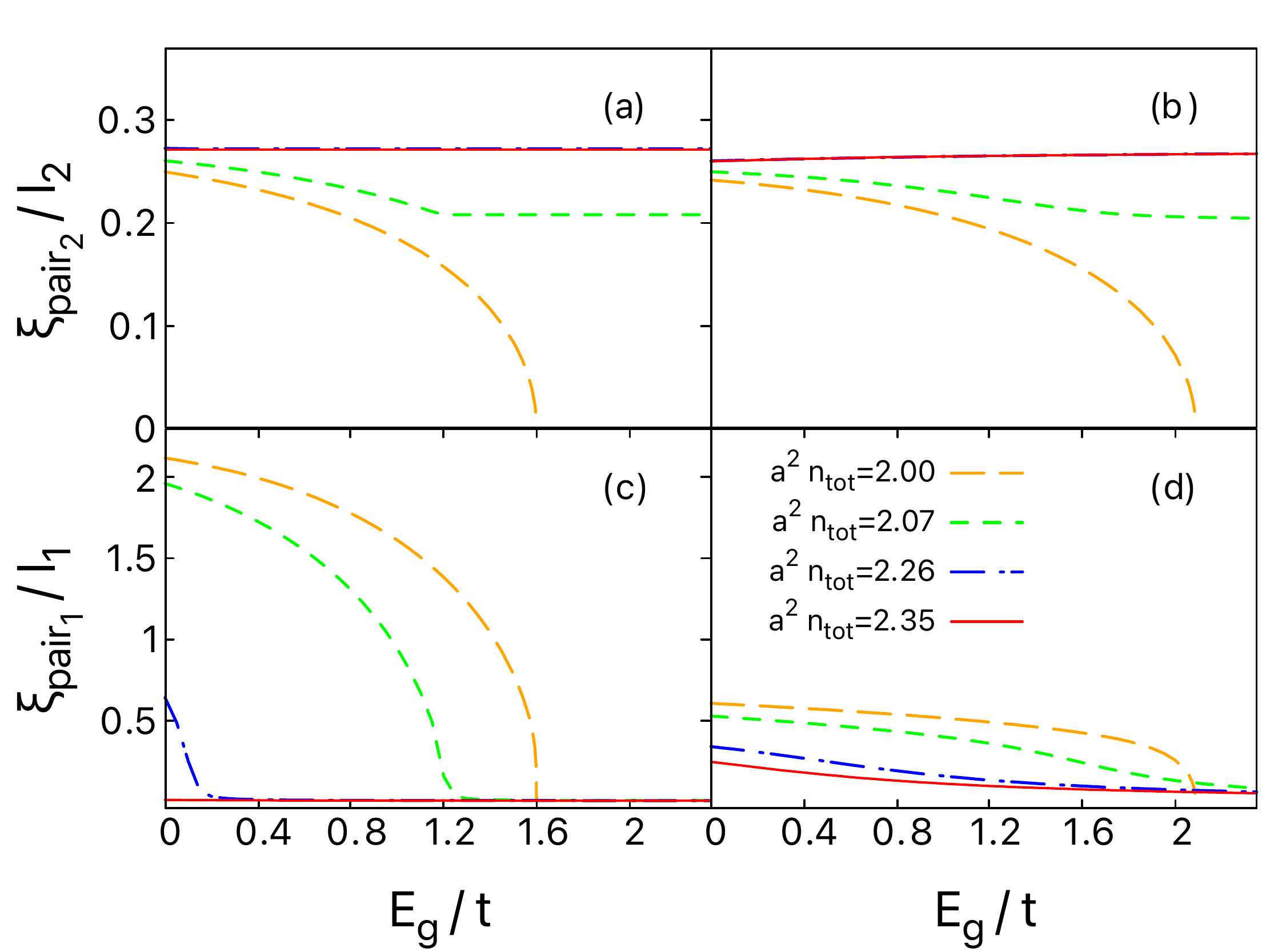}
\caption{Intra-pair coherence length $\xi_{pair 2} / l_2$ for the Cooper pairs of the conduction band (a)-(b) and intra-pair coherence length $\xi_{pair 1} / l_1$ for the Cooper pairs of the valence band (c)-(d) as functions of the band-gap $E_g/t$ for $\omega_0/t=20$. The intra-band couplings are $\lambda_{11}=0.23$ and $\lambda_{22}=0.75$. The pair-exchange couplings are ($\lambda_{12}=\lambda_{21}$): (a),(c) (0.001), (b),(d) (0.1). The intra-pair coherence lengths $\xi_{pair_i} / l_i$ are reported for different $a^2 n_{tot}$.}
\label{fig6}
\end{figure}
\\In Fig. \ref{fig6} the intra-pair coherence length is reported as a function of $E_g$, for different $a^2 n_{tot}$ and for different pair-exchange couplings. Since for low densities and small pair-exchange couplings the valence band condensate is in the BCS regime (\ref{fig6}(a)) when $E_g$ is small, $\xi_{pair_1}$  assumes initially larger values with respect to the average inter-particle distance $l_1$. For larger $E_g$ the system enters the BEC regime and $\xi_{pair_1}$ becomes much smaller than the average inter-particle distance. The valence band condensate goes from the crossover to the BEC regime in a small range of band gap values. This behavior is observed also for larger values of the total density.
\begin{figure}[h]
\hspace{-0.75cm}\includegraphics[width=0.5\textwidth]{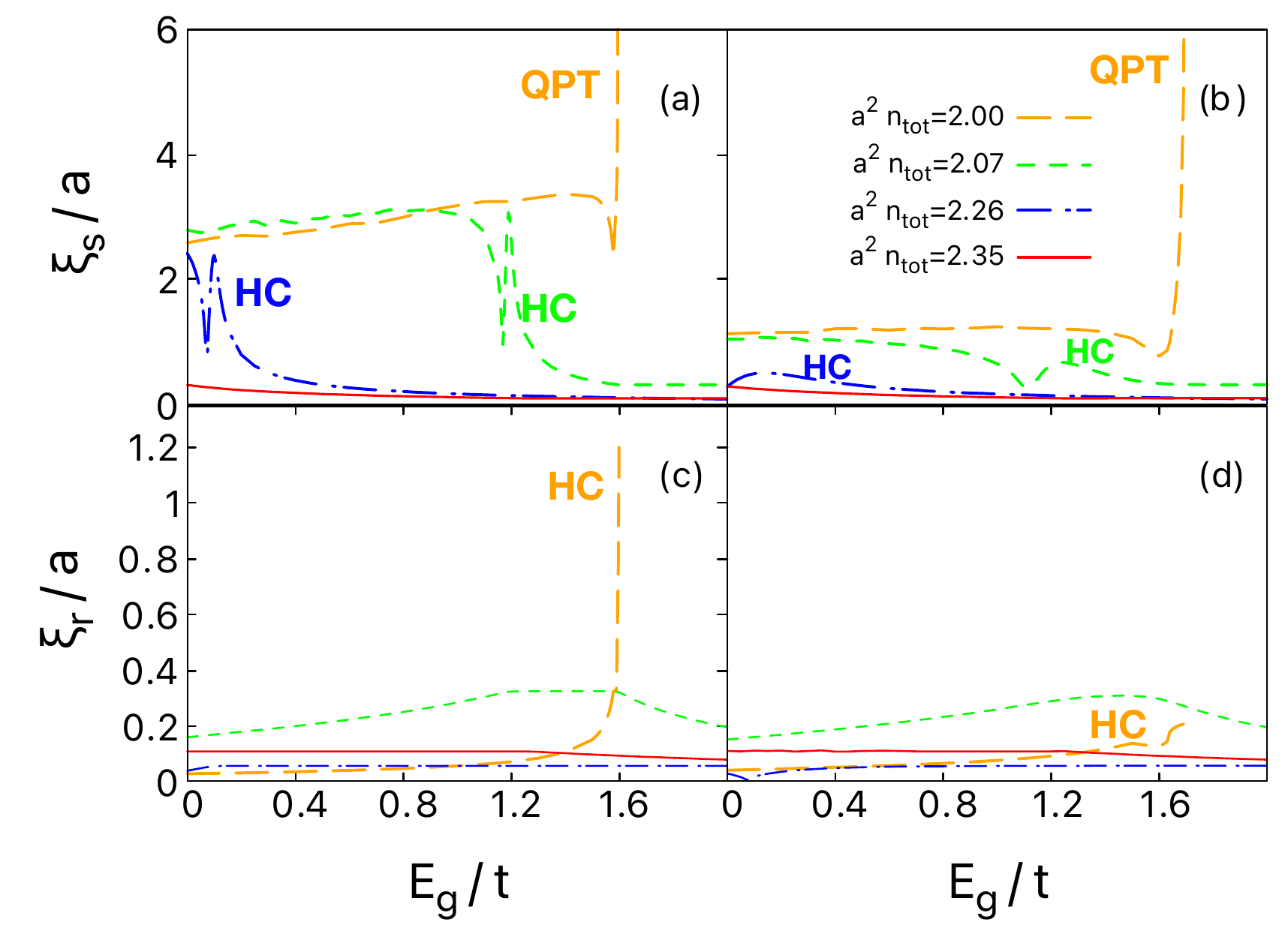}
\caption{Soft $\xi_s$ (a)-(b) and rigid $\xi_r$ (c)-(d) order parameter coherence length, normalized to the lattice constant $a$, as functions of the band-gap $E_g/t$ between the two bands at temperature $T / t=0.00065$ and for $\omega_0 / t=20$. The intra-band couplings are $\lambda_{11}=0.23$ and $\lambda_{22}=0.75$. The pair-exchange couplings are ($\lambda_{12}=\lambda_{21}$): (a),(c) (0.001), (b),(d) (0.03). The coherence lengths $\xi_{s, r}$ are reported for different values of the total density $a^2 n_{tot}$. In the case $a^2 n_{tot}=2.00$ (orange dashed line)  $\xi_r$ has been rescaled by a factor of 7 (c) and 4.5 (d) to make the plot more visible.}
\label{fig7}
\end{figure}
The conduction band instead, due to the strong intra-band coupling retains a small value of the intra-pair coherence length with respect to the the average inter-particle distance $l_2$ for all the considered values of the system density. In this way we found Cooper pairs of different size coexisting in the system for low density and low pair-exchange couplings values, in the regime of small $E_g$. For the zero doping case the intra-pair coherence length is defined only for $E_g<E_g^*$, since in this regime the system is not superconducting and a intra-pair coherence length cannot be defined. The fact that the intra-pair coherence length is approaching zero at the QCP in the BEC regime is different from Ref. \cite{Tajima2019b}, where giant Cooper pairs are found in the vicinity of the QCP in the BCS side. In this case instead, what we have found is equivalent to the finite-density to zero-density QCP of tightly bound molecules. Namely, near the present QCP in the BEC side the pair size is so small that pairs behave as point-like bosons and the system can be described by its bosonic counterpart \cite{Koichiro2022}.\\
In Fig. \ref{fig7} the order parameter coherence coherence length is reported as a function of $E_g$, for different $a^2 n_{tot}$ and for different pair-exchange couplings. In the case $a^2 n_{tot}=2.00$ the soft or critical coherence length $\xi_{s}$ diverges when the band gap reaches the critical value $E_g = E_g^*$, since the system undergoes a quantum phase transition to the insulating state. In the other cases $a^2 n_{tot}\neq2.00$, the soft coherence length $\xi_{s}$ is not diverging, since no quantum phase transition occurs in the system for any $E_g$. In particular, in the cases of $a^2 n_{tot}=2.07$ and $a^2 n_{tot}=2.26$ the soft coherence length $\xi_{s}$ shows a maximum in correspondence of the respective $E_g=E_g^*$, showing its memory about the quantum phase transition of the valence band condensate, which takes place when the pair-exchange interactions are absent. The increase of $\lambda_{12}=\lambda_{21}$ suppresses the maximum, as shown in Figs. \ref{fig7}(a) and (b), since the band-condensates become more coupled.
In the case of $a^2 n_{tot}=2.35$ instead, since the valence band is never superconducting for any $E_g$ when the band-condensates are decoupled, there is no quantum phase transition and no peak.
\begin{figure}[h]
\vspace{1.2cm}\hspace{-0.75cm}\includegraphics[width=0.5\textwidth]{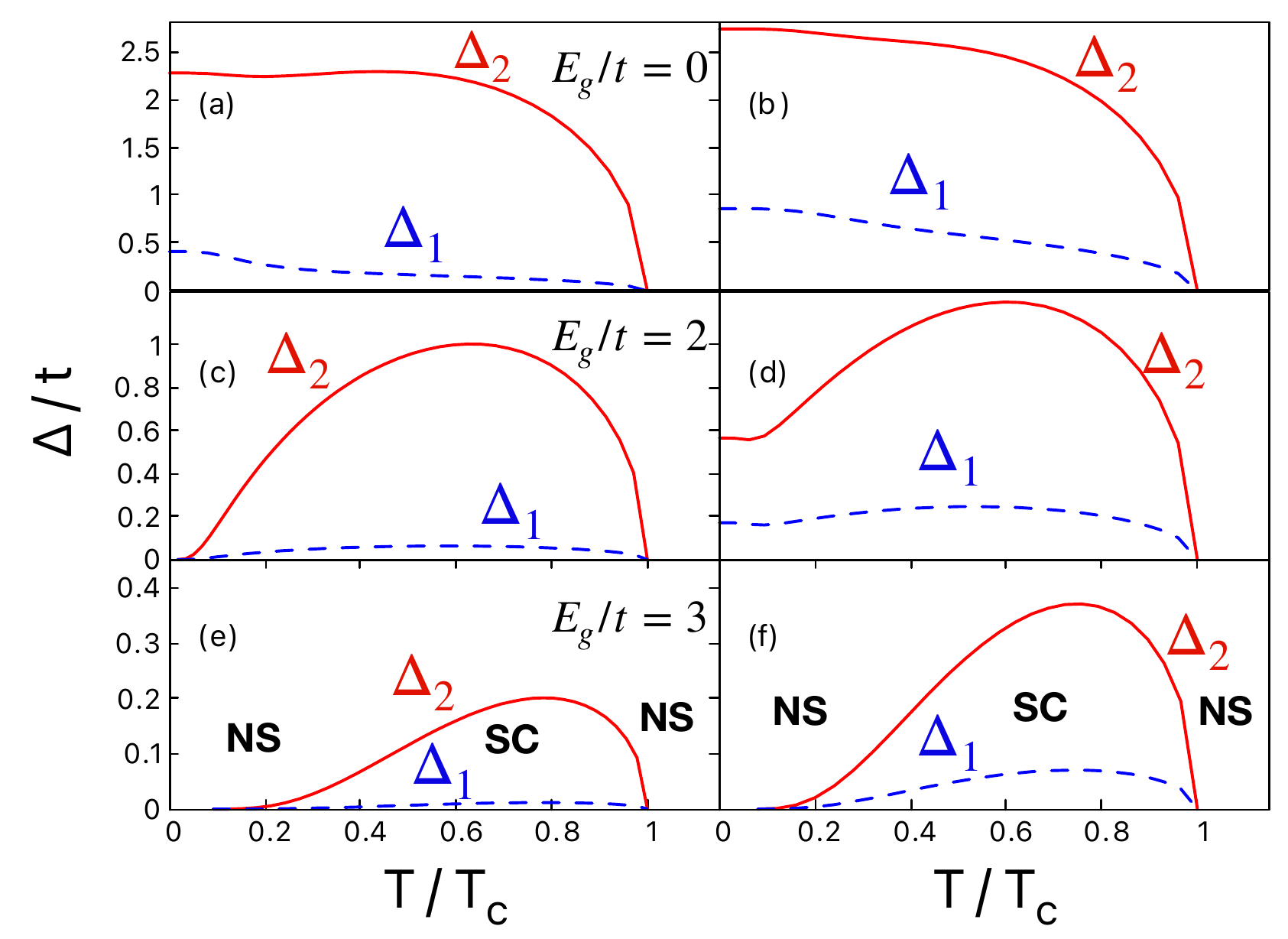}
\caption{Superconducting gaps $\Delta_2 / t$ opening in the conduction band and in the valence band $\Delta_1 / t$ as functions of temperature $T$, normalized with respect to the critical temperature $T_c$, for $a^2 n_{tot}=2.00$. The pair-exchange couplings are ($\lambda_{12}=\lambda_{21}$): (a), (c), (e) (0.03), (b), (d), (f) (0.1).}
\label{fig9}
\end{figure}
The rigid coherence length $\xi_{r}$ instead remains finite for all $E_g$ and for all $a^2 n_{tot}$. Anyway, we find the memory of the quantum phase transition that takes place when the conduction band is empty and the valence band is filled ($a^n n_{tot}=2.00$). In this case in fact, also the conduction band returns to the normal state at  $E_g=E_g^*$. Indeed, for zero pair-exchange couplings, the rigid coherence length $\xi_{r}$ reduces to the coherence length of the conduction band $\xi_2$. Even though for finite pair-exchange coupling the coherence length is non-diverging, it encodes the memory of the quantum phase transition of the conduction band.
Also the maximum value of the rigid coherence length $\xi_{r}$ is suppressed by the increase of $\lambda_{12}=\lambda_{21}$ in this case, as shown in Figs. \ref{fig7}(c) and (d). 
\\We consider now finite temperature effects on the critical energy band gap for the case of no doping. The superconducting gaps as functions of temperature for different band gaps are reported in Fig. \ref{fig9}. The superconducting gaps present a non-monotonic behavior, that is very different from the temperature dependence of the gaps in conventional superconductors. The strong enhancement of $\Delta_2$ at finite temperature is due to the thermal excitation of the electrons from the valence band to the conduction band. This behavior becomes more pronounced for larger $E_g$, especially in the case of Fig. \ref{fig9}(c) in which the system is initially in the normal state for temperatures close to zero, and then becomes superconducting for larger temperatures. This superconducting-normal state reentrant transition that we have found in our two-band system is based on a different mechanism with respect to the reentrant transitions observed in superconductors containing magnetic elements \cite{Eisaki1994} or in granular superconducting systems \cite{Suzuki1983, Lin1984, Welp1988, Chudinov2002}: in the former it is attributed to the competition of magnetic ordering and superconductivity, while in the latter is attributed to tunneling barriers effect, while in our valence-conduction bands system the thermal excitation of electrons from the valence into the conduction band play a crucial role.
In Fig. \ref{fig8} we report the phase diagram $T$ vs $E_g$ for our system. 
\begin{figure}[h]
\hspace{-0.35cm}\includegraphics[width=0.48\textwidth]{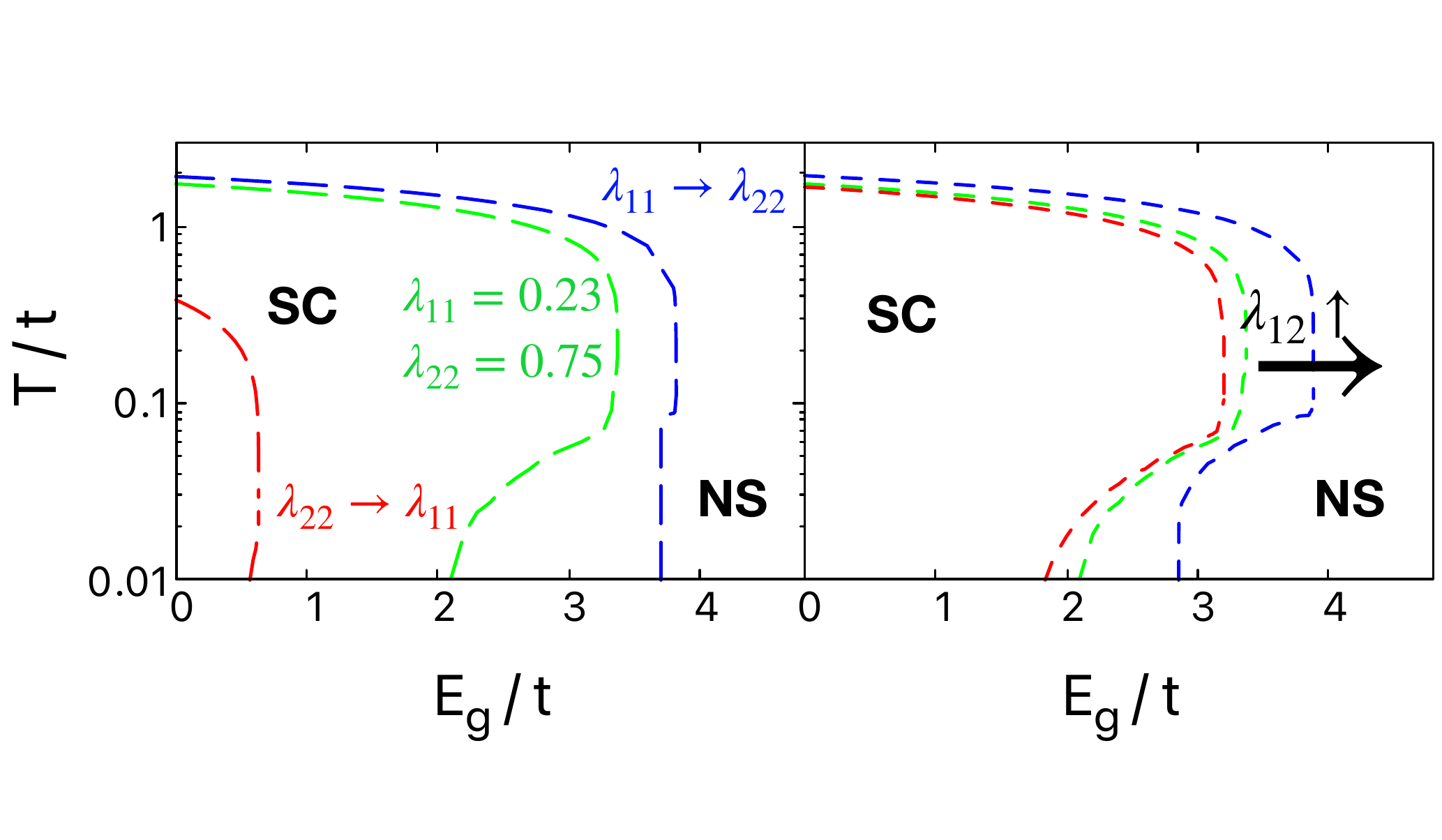}
\caption{Phase diagrams in the temperature versus energy band gap plane, for the zero doping case. In the left panel the red dashed line is for $\lambda_{11}=0.23$, $\lambda_{22}=0.4$, the green dashed line is for $\lambda_{11}=0.23$, $\lambda_{22}=0.75$ and the blue dashed line is for $\lambda_{11}=0.65$, $\lambda_{22}=0.75$. The pair-exchange couplings are the same for all curves, $\lambda_{12}=\lambda_{21}=0.1$. In the right panel the pair-exchange couplings from left to right are: $\lambda_{12}=\lambda_{21}=0.03, 0.1, 0.2$, while the intra-band couplings are $\lambda_{11}=0.23$ and $\lambda_{11}=0.75$. }
\label{fig8}
\end{figure}
In Fig. \ref{fig8} the branch of the phase transition from the superconducting to the normal state corresponding to the reentrant behavior results from
the second solution at lower temperatures of the linearized self-consistent equations for the superconducting gaps. 
From the left panel of Fig. \ref{fig8} it is clear how the reentrant transition is more pronounced when the intra-band couplings are unbalanced ($\lambda_{22} \simeq 3\lambda_{11}$ in the figure), while the reentrance is reduced when the intra-band couplings have similar values. This effect occurs in a less evident manner also when the pair-exchange couplings are increased. Therefore, the most relevant parameter to control the reentrance phenomenon is the intra-band coupling.

\section{Conclusions}
We have studied the superconducting properties of a two-band system of electrons, interacting through a separable attractive potential with a large energy cutoff and multiple pairing channels, at a mean-field level. The superconducting state properties are studied by varying the energy gap between the bands. We have considered different levels of filling for the conduction band, while the valence band is always completely filled. When the band-gap is modified, the density of electrons in the two bands changes, allowing for the occurrence of a density-induced BCS-BEC crossover. When the pair-exchange couplings are small, the condensate in the valence band remains superconducting but with a strongly suppressed superconducting gap $\Delta_1$ for $E_g>E_g^*$. Therefore, in the regime of small pair-exchange coupling, after $E_g^*$, there is only one significant superconducting gap and one significant condensate. Interestingly, in this case the soft coherence length present a peak as a memory of the quantum phase transition that the valence band condensate undergoes in absence of pair exchanges. This peak is more pronounced if the pair-exchange couplings are sufficiently weak and disappears for higher values of the pair-exchange couplings. For higher values of $\lambda_{ij}$, superconductivity in the valence band is sustained by the condensate in the conduction band. Furthermore, in this regime we have found that superconductivity is enhanced in the valence band for increasing doping as long as $E_g<E_g^*$, while for $E_g>E_g^*$ superconductivity is enhanced for lower doping.
We have also found that superconductivity may occur even when no free carriers exist in the conduction band in the normal state at $T = 0$, as soon as the gain in superconducting energy exceeds the cost in producing carriers across the band gap $E_g$. If the binding energy is larger than the energy band-gap, the system becomes unstable under the formation of Cooper pairs and superconductivity emerges. However, there exists a critical value of the energy band gap $E_g^*$ in correspondence of which the process of creating Cooper pairs is not energetically favorable anymore, at this point a quantum phase transition occurs. This quantum phase transition is confirmed by the soft coherence length, which is diverging in correspondence of the critical band gap $E_g=E_g^*$. Thus, the ground state is superconducting if $E_g < E_g^*$, insulating if $E_g > E_g^*$. At finite temperature, the value of $E_g^*$ is larger than its zero temperature value, because the electrons are thermally excited from the valence band. This situation is responsible for the non-monotonic behavior of the superconducting gap opening in the conduction band, which is enhanced at low temperatures because of the electrons that jump from the valence band into the conduction band due to thermal excitation. When there is a finite doping in the system, the sharp phase transition becomes a smooth crossover and superconductivity extends for all $E_g$. In this case, for $E_g > E_g^*$ the valence band contributes very weakly to the superconducting state, since the hole density becomes almost zero in this regime. 
\\To conclude, we have found that the system explores different regimes of the BCS-BEC crossover by tuning the energy band-gap and the total density. The valence-band condensate spans the entire BCS-BEC crossover for low enough density by varying the band-gap $E_g$. For larger values of the total density, the condensate of the valence band is very dilute and results in the BEC regime for any $E_g$. The condensate of the conduction band instead resides in the BEC side of the crossover or completely inside the BEC regime, due to the strength of the intra-band coupling of electrons in the conduction band. This picture of the BCS-BEC crossover for the system has been found by analyzing the consistent behavior of the chemical potential, the condensate fractions and the coherence lengths. Finally, in the case of zero doping and at finite temperature, an interesting new type of reentrant superconducting to normal state transition has been numerically discovered for unbalanced intra-band couplings, showing that in this configuration superconductivity is assisted instead of being suppressed by increasing temperature. This happens because the electrons in the valence band are able to jump into the conduction band even for larger values of the zero temperature critical band gap, due to thermal excitation, making the superconducting state available for a wider range of $E_g$ when the temperature is higher.
\section{Acknowledgments}
We are grateful to Tiago Saraiva (HSE-Moscow) and Hiroyuki Tajima (University of Tokyo) for interesting discussions and a critical reading of the manuscript.  
G. M. acknowledges INFN for financial support of his Ph.D. grant. This work has been partially supported by PNRR MUR project PE0000023-NQSTI.


\begin{thebibliography}{1}

\bibitem{Milosevic2015}
Milorad V. Milošević and Andrea Perali, Emergent phenomena in multicomponent superconductivity: an introduction to the focus issue, Supercond. Sci. Technol. $\textbf{28}$, 060201 (2015).

\bibitem{Eagles1969}
D. M. Eagles, Possible Pairing without Superconductivity at Low Carrier Concentrations in Bulk and Thin-Film Superconducting Semiconductors, Phys. Rev. $\textbf{186}$, 456 (1969).

\bibitem{Leggett1980}
A. J. Leggett, in \textit{Modern Trends in the Theory of Condensed Matter}, edited by A. Pekelski and J. Przystawa (Springer-Verlag, Berlin, 1980), p. 13.

\bibitem{Chen2005}
Q. Chen, J. Stajic, S. Tan, and K. Levin, BCS–BEC crossover: From high temperature superconductors to ultracold superfluids, Phys. Rep. $\textbf{412}$, 1 (2005).

\bibitem{Strinati2018}
G. C. Strinati, P. Pieri, G. Röpke, P. Schuck, and M. Urban, The BCS–BEC crossover: From ultra-cold Fermi gases to nuclear systems, Phys. Rep. $\textbf{738}$, 1 (2018).

\bibitem{Shanenko2010}
A. A. Shanenko, M. D. Croitoru, A. Vagov, and F. M. Peeters, Giant drop in the Bardeen-Cooper-Schrieffer coherence length induced by quantum size effects in superconducting nanowires, Phys. Rev. B $\textbf{82}$, 104524 (2010).

\bibitem{Chen2012}
Y. Chen, A. A. Shanenko, A. Perali, and F. M. Peeters, Superconducting nanofilms: molecule-like pairing induced by quantum confinement, J. Phys.: Condens. Matter $\textbf{24}$, 185701 (2012).

\bibitem{Innocenti2010}
D. Innocenti, N. Poccia, A. Ricci, A. Valletta, S. Caprara, A. Perali, and A. Bianconi, Resonant and crossover phenomena in a multiband superconductor: Tuning the chemical potential near a band edge, Phys. Rev. B $\textbf{82}$, 184528 (2010).

\bibitem{Mazziotti2017}
M. V. Mazziotti, A. Valletta, G. Campi, D. Innocenti, A. Perali, and A. Bianconi, Possible Fano resonance for high-Tc multi-gap superconductivity in p-Terphenyl doped by K at the Lifshitz transition, Eur. Phys. Lett. $\textbf{118}$, 37003 (2017).

\bibitem{Salasnich2019}
L. Salasnich, A. A. Shanenko, A. Vagov, J. Albino Aguiar, and A. Perali, Screening of pair fluctuations in superconductors with coupled shallow and deep bands: A route to higher-temperature superconductivity, Phys. Rev. B $\textbf{100}$, 064510 (2019).

\bibitem{Tajima2019a}
H. Tajima, Y. Yerin, A. Perali, P. Pieri, Enhanced critical temperature, pairing fluctuation effects, and BCS-BEC crossover in a two-band Fermi gas, Phys. Rev. B  $\textbf{99}$, 180503(R) (2019).

\bibitem{Tajima2020}
H. Tajima, Y. Yerin, P. Pieri, A. Perali, Mechanisms of screening or enhancing the pseudogap throughout the two-band Bardeen-Cooper-Schrieffer to Bose-Einstein condensate crossover, Phys. Rev. B $\textbf{102}$, 220504(R) (2020).

\bibitem{Saraiva2020}
T. T. Saraiva, P. J. F. Cavalcanti, A. Vagov, A. S. Vasenko, A. Perali, L. Dell'Anna, and A. A. Shanenko, Multiband Material with a Quasi-1D Band as a Robust High-Temperature Superconductor, Phys. Rev. Lett. $\textbf{125}$, 217003 (2020). 

\bibitem{Saraiva2021}
T. T. Saraiva, L. I. Baturina, and A. A. Shanenko, Robust Superconductivity in Quasi-one-dimensional Multiband Materials, The Journal of Physical Chemistry Letters $\textbf{12}$, 11604 (2021). 

\bibitem{Saraiva2022}
 A. A. Shanenko, T. T. Saraiva, A. Vagov, A. S. Vasenko, and A. Perali, Suppression of fluctuations in a two-band superconductor with a quasi-one-dimensional band, Phys. Rev. B $\textbf{105}$, 214527 (2022).

\bibitem{Gabovich2009}
Alexander M. Gabovich, and Alexander I. Voitenko, Model for the coexistence of d-wave superconducting and charge-density-wave order parameters
in high-temperature cuprate superconductors, Phys. Rev. B $\textbf{80}$, 224501 (2009).


\bibitem{Gabovich2010}
Alexander M. Gabovich and Alexander I. Voitenko, Coexistence of charge density waves and d-wave superconductivity in cuprates. Sharing of the Fermi surface, Z. Kristallogr. $\textbf{225}$, 492 (2010).

\bibitem{Arpaia2019}
R. Arpaia, S. Caprara, R. Fumagalli, G. De Vecchi, Y. Y. Peng, E. Andersson, D. Betto, G. M. De Luca, N. B. Brookes, F. Lombardi, M. Salluzzo, L. Braicovich, C. Di Castro, M. Grilli, and G. Ghiringhelli, Dynamical charge density fluctuations pervading the phase diagram of a Cu-based high-Tc superconductor, Science $\textbf{365}$, 906 (2019).

\bibitem{Perali1996}
A. Perali, C. Castellani, C. Di Castro, and M. Grilli, d-wave superconductivity near charge instabilities, Phys. Rev. B $\textbf{54}$, 16216 (1996).

\bibitem{Rossnagel2011}
Rossnagel K., On the origin of charge density waves in select layered transition-metal dichalcogenides, J. Phys. Condens. Matter $\textbf{23}$, 213001 (2011).

\bibitem{Neto2001}
 A. H. Castro Neto, Charge Density Wave, Superconductivity, and Anomalous Metallic Behavior in 2D Transition Metal Dichalcogenides, Phys. Rev. Lett. $\textbf{86}$, 4382 (2001).

\bibitem{Kiss2007}
T. Kiss, T. Yokoya, A. Chainani, S. Shin, T. Hanaguri, M. Nohara,
and H. Takagi, Charge-order-maximized momentum-dependent superconductivity, Nat. Phys. $\textbf{3}$, 720 (2007).

\bibitem{Calandra2009}
M. Calandra, I. I. Mazin, and F. Mauri, Effect of dimensionality on the charge-density wave in few-layer 2H-NbSe$_2$, Phys. Rev. B $\textbf{80}$, 241108(R) (2009).

\bibitem{Ge2012}
Y. Ge, and A. Y. Liu, Effect of dimensionality and spin-orbit coupling on charge-density-wave transition in 2H-TaSe, Phys. Rev. B $\textbf{86}$, 104101 (2012).

\bibitem{Ugeda2016}
Ugeda M. M. et al., Characterization of collective ground states in single-layer NbSe$_2$,  Nat. Phys. $\textbf{12}$, 92 (2016).

\bibitem{Cao2015}
Cao Y. et al., Quality Heterostructures from Two-Dimensional Crystals Unstable in Air by Their Assembly in Inert Atmosphere, Nano Lett. $\textbf{15}$, 4914 (2015).

\bibitem{Lian2019}
Chao-Sheng Lian, Christoph Heil, Xiaoyu Liu, Chen Si, Feliciano Giustino, and Wenhui Duan, Coexistence of Superconductivity with Enhanced Charge-Density Wave Order in the Two-Dimensional Limit of TaSe, J. Phys. Chem. Lett. $\textbf{10}$, 4076 (2019).

\bibitem{Liu2015}
Ge J. F., Liu Z. L., Liu C. et al., Superconductivity above $100$ K in single-layer FeSe films on doped SrTiO$_3$, Nature Mater $\textbf{14}$, 285 (2015).

\bibitem{Zhou2022} 
Zhou H., Holleis L., Saito Y., Cohen L., Huynh W., Patterson C.L., Yang F., Taniguchi T., Watanabe K., and Young A.F., Isospin magnetism and spin-polarized superconductivity in Bernal bilayer graphene, Science $\textbf{375}$, 774 (2022).  

\bibitem{Pantaleon2022} 
Pierre A. Pantaleon, Alejandro Jimeno-Pozo, Hector Sainz-Cruz, Tommaso Cea, Vo Tien Phong, and Francisco Guinea, Superconductivity and correlated phases in bilayer, trilayer graphene and related structures, 	arXiv:2211.02880 (2022).

\bibitem{Zhang2022} 
Yiran Zhang, Robert Polski, Alex Thomson, Étienne Lantagne-Hurtubise, Cyprian Lewandowski, Haoxin Zhou, Kenji Watanabe, Takashi Taniguchi, Jason Alicea, and Stevan Nadj-Perge, Spin-Orbit Enhanced Superconductivity in Bernal Bilayer Graphene, arXiv:2205.05087 (2022).


\bibitem{Conti2017}
S. Conti, A. Perali, F. M. Peeters, and D. Neilson, Multicomponent Electron-Hole Superfluidity and the BCS-BEC Crossover in Double Bilayer Graphene, Phys. Rev. Lett. $\textbf{119}$, 257002 (2017).


\bibitem{Nozieres1999}
P. Nozieres, and F. Pistolesi, From semiconductors to superconductors: a simple model for pseudogaps, Eur. Phys. J. B $\textbf{10}$, 649 (1999).

\bibitem{Tajima2019b} Y. Yerin, H. Tajima, P. Pieri, A. Perali, Coexistence of giant Cooper pairs with a bosonic condensate and anomalous behavior of energy gaps in the BCS-BEC crossover of a two-band superfluid Fermi gas, 
Phys. Rev. B $\textbf{100}$, 104528 (2019).

\bibitem{Andrenacci1999}
N. Andrenacci, A. Perali, P. Pieri, and G.C. Strinati, Density-induced BCS to Bose-Einstein crossover, Phys. Rev. B $\textbf{60}$, 12410 (1999).

\bibitem{Carlos2022}
Yue-Ran Shi, Wei Zhang, and C. A. R. Sá de Melo, The evolution from BCS to Bose pairing in two-band superfluids: Quantum phase transitions and crossovers by tuning band offset and interactions, EPL $\textbf{139}$, 36004 (2022).

\bibitem{Ord2012}
Teet Ord, Kullike Rago, and Artjom Vargunin, Critical and non-critical coherence lengths in a two-band superconductor, Journal of Superconductivity and Novel Magnetism $\textbf{25}$, 1351 (2012).

\bibitem{Guidini2014} A. Guidini, A. Perali, Band-edge BCS - BEC crossover in a two-band superconductor: Physical properties and detection parameters, Supercond. Sci. and Technol. $\textbf{27}$, 124002 (2014). 

\bibitem{Paredes2020}
A. A. Vargas-Paredes, A. A. Shanenko, A. Vagov, M. V. Milošević, and A. Perali, Crossband versus intraband pairing in superconductors: Signatures and consequences of the interplay,
Phys.Rev.B $\textbf{101}$, 094516 (2020).

\bibitem{Koichiro2022}
K. Furutani, A. Perali, and L. Salasnich., Berezinskii-Kosterlitz-Thouless phase transition with Rabi coupled bosons, arXiv:2210.10866, (2022).


\bibitem{Eisaki1994}
H. Eisaki, H. Takagi, R.J. Cava, B. Batlogg, 3.J. Krajewski, W.F. Perk, Jr., K. Mizuhashi, J.O. Lee, and S. Uchida, Phys. Rev. B $\textbf{50}$, 647 (1994).

\bibitem{Suzuki1983}
T. Suzuki, T. Tsuboi, H. Takaki, T. Nizusaki, and T. Kusumoto, J. Phys. Soc. Jpn $\textbf{52}$, 981 (1983).

\bibitem{Lin1984}
T.H. Lin, X.Y. Shao, M.K. Wu, P.H. Hor, X.C..]in, and C.W. Chu, Phys. Rev. B $\textbf{29}$, 1493 (1984).

\bibitem{Welp1988}
U.Welp, W.K. Kwok, G.W. Crabtree, H. Claus, K.G. Vandervoort, B. Dabrowski, A.W. Mitchell, D.R. Richards, D.T. Mark, and D.G. Hinks, Physica C $\textbf{156}$, 27 (1988).

\bibitem{Chudinov2002}
S. M. Chudinov, G. Mancini, M. Minestrini, R. Natali, S. Stizza, and A. Bozhko, J. Phys. Condens. Matter $\textbf{14}$, 193 (2002).

\end{thebibliography}
\end{document}